\documentclass{article}%
\usepackage{amsmath}
\usepackage{amsfonts}
\usepackage{amssymb}
\usepackage{graphicx}%
\begin{document}

\author{Sawa Manoff\\\textit{Bulgarian Academy of Sciences}\\\textit{Institute for Nuclear Research and Nuclear Energy}\\\textit{Department of Theoretical Physics}\\\textit{Blvd. Tzarigradsko Chaussee 72}\\\textit{1784 Sofia - Bulgaria}}
\date{E-mail address: smanov@inrne.bas.bg}
\title{Standard emitters (clocks) and calibrated standard emitters (clocks) in spaces
with affine connections and metrics}
\maketitle

\begin{abstract}
It is shown that the general belief that the frequency and the absolute value
of the velocity of periodic signals sent by a standard emitter do not change
on the world line of the emitter needs to be revised and new conditions for
the existence of a calibrated standard emitter should be taken into account.
The definitions of a standard clock and of a calibrated standard clock are
introduced in a space with affine connections and metrics. The variation of
the velocity and of the frequency of a standard clock could be compared with
the constant velocity and the constant frequency of a (calibrated) standard
clock along the world line of the observer. This calibrated standard clock is
transported by means of a generalized Fermi-Walker transport along the same
world line of the observer. Some remarks about the synchronization of standard
clocks in spaces with affine connections and metrics are given.


\textit{Short title}: Standard clocks and calibrated standard clocks in
$(\overline{L}_{n}$,$g)$-spaces


\end{abstract}

\section{Introduction}

\subsection{General remarks.}

1. Modern problems of relativistic astrophysics as well as of relativistic
physics are related to the propagation of signals in space or in space-time.
The basis of experimental data received as results of observations of the
Doppler effect or of the Hubble effect gives rise to considerations about the
theoretical status of effects related to detection of signals from emitters
moving relatively to observers carrying detectors in their laboratories.
Nevertheless, in the last decades, there is no essential evolution of the
theoretical models related to new descriptions of the Doppler and Hubble
effects corresponding to the recent development of new mathematical models for
the space-time. In the astronomy and astrophysics standard theoretical schemes
for measuring velocities are used related to classical mechanics and / or
special and general relativity \cite{Lindegren}, \cite{Eva-Maria}.

2. The incoming periodic signals sent by an emitter moving relatively to an
observer (detector) are compared with periodic signals of an emitter (standard
emitter) lying at rest with the observer. On this basis, the change of
frequency and velocity of the incoming periodic signals leads to conclusions
about velocities and accelerations of objects moving with respect to the
observer. By that \textit{it is assumed} that the periodic signals of an
emitter lying at rest with the observer have constant frequency and constant
absolute value of their velocity along the world line of the observer. \ This
assumption is based on the fact that there are different emitters constructed
in a way (e.g. by the use of the s.c. method of automatic frequency
modulation) that their frequencies remain constant in time. \ In general, it
is worth to be investigated how the kinematic characteristics (shear,
rotation, and expansion velocities and accelerations) of the motion of the
observer (respectively of the standard emitter) could influence the frequency
and the absolute value of the velocity of the signals emitted by a standard
emitter considered in more comprehensive models of space-time such as spaces
with affine connections and metrics [$(\overline{L}_{n},g)$-spaces].

3. In recent years, spaces with affine connections and metrics have received
some interest related to the possibility of using mathematical models of
space-time different from (pseudo-) Riemannian spaces without torsion ($V_{n}%
$-spaces) or with torsion ($U_{n}$-spaces). It has been shown \cite{Manoff-1},
\cite{Manoff-1a}, \cite{Manoff-2} that every differentiable manifold $M$
($dimM=n$) with affine connections and metrics [$(\overline{L}_{n},g)$-spaces]
\cite{Manoff-1} could be used as models of space-time for the following reasons:

- The equivalence principle (related to the vanishing of the components of an
affine connection at a point or on a curve) holds in $(\overline{L}_{n}%
,g)$-spaces \cite{Iliev-1}$\div$\cite{Iliev-1b}, \cite{Hartley}.

- $(\overline{L}_{n},g)$-spaces have structures similar to these in (pseudo)
Riemannian spaces without torsion [$V_{n}$-spaces] allowing for description of
dynamic systems and the gravitational interaction \cite{Manoff-2}.

- Fermi-Walker transports and conformal transports exist in $(\overline{L}%
_{n},g)$-spaces as generalizations of these types of transports in $V_{n}%
$-spaces \cite{Manoff-4a}, \cite{Manoff-4b}.

- A Lorentz basis and a light cone could not be deformed in $(\overline{L}%
_{n},g)$-spaces as it is the case in $V_{n}$-spaces.

- All kinematic characteristics related to the notions of relative velocity
and of relative acceleration could be worked out in $(\overline{L}_{n}%
,g)$-spaces without changing their physical interpretations in $V_{n}$-spaces
\cite{Manoff-2}, \cite{Manoff-7}, \cite{Manoff-8}.

- $(\overline{L}_{n},g)$-spaces include all types of spaces with affine
connections and metrics used until now as models of space-time.

On this basis, many of the differential-geometric construction used in the
Einstein theory of gravitation (ETG) in $V_{4}$-spaces could be generalized
for the cases of spaces with one affine connection and metrics [$(L_{n}%
,g)$-spaces] and in spaces with affine connections and metrics [$(\overline
{L}_{n},g)$-spaces]. Bearing in mind this background a question arises about
possible physical applications and interpretation of mathematical
constructions from ETG generalized for $(\overline{L}_{n},g)$-spaces. \ The
theory of $(\overline{L}_{n},g)$-spaces is worked out in details in
\cite{Manoff-1}. Brief review of the properties of $(\overline{L}_{n}%
,g)$-spaces is given in \cite{Manoff-1a} or in \cite{Manoff-7}.

4. It is well known that every classical field theory over spaces with affine
connections and metrics could be considered as a theory of continuum media in
these spaces \cite{Schmutzer} $\div$ \cite{Manoff-8b}. On this ground, notions
of the continuous media mechanics (such as deformation velocity and
acceleration, shear velocity and acceleration, rotation velocity and
acceleration, expansion velocity and acceleration) have been used as invariant
characteristics for spaces admitting vector fields with special kinematic characteristics.

5. If a $(\overline{L}_{n},g)$-space \ could be used as a model of space or of
space-time the questions arise

\begin{itemize}
\item how signals propagate in a space-time described by a $(\overline{L}%
_{n},g)$-space and

\item how signals could be influenced by the kinematic characteristics of the
motion of an observer.
\end{itemize}

The answer of the first question is given in details in \cite{Manoff-8c}. The
answer of the second question is the subject of this paper.

\textit{Remark.} In a previous paper \cite{Manoff-8d} the variation of the
velocity and the frequency of a periodic signal along the world line of the
emitter has been considered without the explicit introduction of the notion of
a calibrated standard emitter (clock). This fact could generate the questions
with respect to what the variation of the velocity and the frequency of a
standard clock could be compared and how the proper time of the frame of
reference of the observer could be introduced.

6. In this paper the variation of the absolute value of the velocity and the
frequency of signals sent by a standard emitter are considered in the proper
frame of reference of the emitter. In Section 2 the variation of the velocity
of a periodic signal along the world line of a standard emitter is considered
as well as the variation of the frequency of a periodic signal along the world
line of a standard emitter is determined. The variation of the velocity and
the frequency of periodic signals of a standard oscillator (clock) are also
found. In Section 3 the calibrated standard clocks are considered and the
conditions for their existence along the world line of an observer are found.
A standard emitter with constant characteristics of its emitted signals (their
absolute value of velocity and their frequency) is called calibrated standard
emitter. The synchronization of calibrated standard clocks on different world
lines is discussed. Some concluding remarks comprise the Section 4. It is
shown that \textit{the general belief that the frequency and the absolute
value of the velocity of periodic signals sent by a standard emitter does not
change on the world line of the emitter needs to be revised and new conditions
for existence of calibrated standard emitter should be taken into account}.

\subsection{Abbreviations, definitions, and symbols}

1. In the further considerations in this paper we will use the following
abbreviations, definitions and symbols:

:= means by definition.

The middle point "$\cdot$" is used as a symbol for standard multiplication in
the field of real (or complex) numbers, e.g. $a\cdot b\in\mathbf{R}$.

The lower point "$.$" is used as a symbol for symmetric tensor product, e.g.
$u.v=\frac{1}{2}\cdot(u\otimes v+v\otimes u)$.

The symbol "$\wedge$" is used as symbol for \ a wedge product, e.g. $u\wedge
v=\frac{1}{2}\cdot(u\otimes v-v\otimes u)$.

$M$ is a symbol for a differentiable manifold with $\dim M=n$. $T(M):=\cup
_{x\in M}T_{x}(M)$ and $T^{\ast}(M):=\cup_{x\in M}T_{x}^{\ast}(M)$ are the
tangent and the cotangent spaces at $M$ respectively. $T_{x}(M)$ and
$T_{x}^{\ast}(M)$ are the tangent and the cotangent spaces at a point $x\in$
$M$ respectively.

\textit{Remark. }The notion of tangent space $T(M)$ over a manifold $M$ could
be considered as an element of the notion of tangent bundle $[T(M)$, $M$,
$\pi:T(M)\rightarrow M]$ over a manifold $M$. A vector field in the tangent
space $T(M)$ over the manifold $M$ could be considered as a section in the
tangent space $T(M)$.

$(\overline{L}_{n},g)$-spaces are spaces with contravariant and covariant
affine connections and metrics whose components \textit{differ not only by
sign}. In such type of spaces the non-canonical contraction operator $S$ acts
on a contravariant basic vector field $e_{j}$ (or $\partial_{j}$)$\,\in
\{e_{j}$ (or $\partial_{j}$)$\}\subset T(M)$ and on a covariant \ basic vector
field $e^{i}\,$\ (or $dx^{i}$) $\in\{e^{i}$ (or $dx^{i}$)$\}\subset T^{\ast
}(M)$ in the form
\begin{align*}
S  &  :(e^{i},e_{j})\longrightarrow S(e^{i},e_{j}):=S(e_{j},e^{i}%
):=f^{i}\,_{j}\text{ , \ \ \ \ \ \ }\\
\text{\ \ \ \ }f^{i}\,_{j}  &  \in C^{r}(M)\text{ , \ \ \ \ \ \ \ \ }%
r\geqq2\text{ , \ \ \ \ \ \ }\det(f^{i}\,_{j})\neq0\text{, \ \ \ \ \ \ }\\
\exists\text{\ \ \ }f_{i}\,^{k}  &  \in C^{r}(M)\text{, \ \ \ \ \ \ \ \ \ }%
r\geqq2\text{ }:\text{\ \ \ \ \ \ \ \ }f^{i}\,_{j}\cdot f_{i}\,^{k}:=g_{j}%
^{k}\text{, \ \ \ \ \ \ \ \ \ \ }%
\end{align*}

In these spaces, for example, $g(u)=g_{ik}\cdot f^{k}\,_{j}\cdot u^{j}\cdot
dx^{i}:=g_{i\overline{j}}\cdot u^{j}\cdot dx^{i}=g_{ij}\cdot u^{\overline{j}%
}\cdot dx^{i}:=u_{i}\cdot dx^{i}$, $g(u,u)=g_{kl}\cdot f^{k}\,_{i}\cdot
f^{l}\,_{j}\cdot u^{i}\cdot u^{j}:=g_{\overline{i}\overline{j}}\cdot
u^{i}\cdot u^{j}=g_{ij}\cdot u^{\overline{i}}\cdot u^{\overline{j}}=u_{j}\cdot
u^{\overline{j}}:=u_{\overline{i}}\cdot u^{i}$, $g^{\overline{i}\overline{j}%
}\cdot g_{jk}=\delta_{k}^{i}=g_{k}^{i}$, $g_{\overline{i}\overline{k}}%
.g^{kj}=g_{i}^{j}$. The components $\delta_{j}^{i}:=g_{j}^{i}$ ($\mid=0$ for
$i\neq j$ and $\mid=1$ for $i=j$) are the components of the Kronecker tensor
$Kr:=g_{j}^{i}\cdot\partial_{i}\otimes dx^{j}$.

$(L_{n},g)$-spaces are spaces with contravariant and covariant affine
connections and metrics whose components \textit{differ only by sign}. In such
type of spaces the canonical contraction operator $S:=C$ acts on a
contravariant basic vector field $e_{j}$ (or $\partial_{j}$)$\,\in\{e_{j}$ (or
$\partial_{j}$)$\}\subset T(M)$ and on a covariant \ basic vector field
$e^{i}\,$\ (or $dx^{i}$) $\in\{e^{i}$ (or $dx^{i}$)$\}\subset T^{\ast}(M)$ in
the form
\[
C:(e^{i},e_{j})\longrightarrow C(e^{i},e_{j}):=C(e_{j},e^{i}):=\delta_{j}%
^{i}:=g_{j}^{i}\text{ .}%
\]

In these spaces, for example, $g(u)=g_{ik}\cdot g_{j}^{k}\cdot u^{j}\cdot
dx^{i}:=g_{ij}\cdot u^{j}\cdot dx^{i}=u_{i}\cdot dx^{i}$, $g(u,u)=g_{kl}\cdot
g_{i}^{k}\cdot g_{j}^{l}\cdot u^{i}\cdot u^{j}:=g_{ij}\cdot u^{i}\cdot
u^{j}=u_{i}\cdot u^{i}$.

\textit{Remark}. All results found for $(\overline{L}_{n},g)$-spaces could be
specialized for $(L_{n},g)$-spaces as well as for all other special cases of
$(L_{n},g)$-spaces, used until now as models of space-time, by omitting the
bars above or under the indices.

$\nabla_{u}$ is the covariant differential operator acting on the elements of
the tensor algebra $\mathcal{T}$ over $M$. The action of \ $\nabla_{u}$ is
called covariant differentiation (covariant transport) along a contravariant
vector field $u$, for instance,
\begin{equation}
\nabla_{u}v:=v_{\;;j}^{i}\cdot u^{j}\cdot\partial_{i}=(v^{i}\,_{,j}%
+\Gamma_{kj}^{i}\cdot v^{k})\cdot u^{j}\cdot\partial_{i}\text{ ,
\ \ \ \ \ }v\in T(M)\text{ ,} \label{0.1}%
\end{equation}
where $v^{i}\,_{,j}:=\partial v^{i}/\partial x^{j}$ and $\Gamma_{jk}^{i}$ are
the components of the contravariant affine connection $\Gamma$ in a
contravariant co-ordinate basis $\{\partial_{i}\}$. The result $\nabla_{u}v$
of the action of $\nabla_{u}$ on a tensor field $v\in\otimes_{l}^{k}(M)$ is
called covariant derivative of $v$ along $u$. For covariant vectors and tensor
fields an analogous relation holds, for instance,
\begin{equation}
\nabla_{u}w=w_{i;j}\cdot u^{j}\cdot dx^{i}=(w_{i,j}+P_{ij}^{l}.w_{l})\cdot
u^{j}\cdot.dx^{i}\text{ \ , \ \ \ }w\in T^{\ast}(M)\text{ .} \label{0.2}%
\end{equation}
where $P_{ij}^{l}$ are the components of the covariant affine connection $P$
in a covariant co-ordinate basis $\{dx^{i}\}$. For $(L_{n},g)$, $U_{n}$, and
$V_{n}$-spaces $P_{ij}^{l}=-\Gamma_{ij}^{l}$.

For $(\overline{L}_{n},g)$-spaces, where $S:(e^{i},e_{j})\longrightarrow
S(e^{i},e_{j}):=S(e_{j},e^{i}):=f^{i}\,_{j}$, the components $\Gamma_{jk}^{i}$
of the contravariant affine connection $\Gamma$ and the components $P_{jk}%
^{i}$ of the covariant affine connection $P$ should obey the condition%
\[
f^{i}~_{j,k}=\Gamma_{jk}^{l}\cdot f^{i}~_{l}+P_{lk}^{i}\cdot f^{l}~_{j}\text{
\ \ \ \ \ \ \ \ , \ \ \ }f^{i}~_{j,k}=\partial_{k}f^{i}~_{j}=\frac{\partial
f^{i}~_{j}}{\partial x^{k}}\text{ \ \ \ \ .\ \ \ \ \ }%
\]

$\pounds _{u}$ is the Lie differential operator \cite{Manoff-1} acting on the
elements of the tensor algebra $\mathcal{T}$ over $M$. The action of
\ $\pounds _{u}$ is called dragging-along a contravariant vector field $u$.
The result $\pounds _{u}v$ of the action of $\pounds _{u}$ on a tensor field
$v$ is called Lie derivative of $v$ along $u$.

2. Let us now recall the physical interpretations of some of the considered
mathematical structures.

\textit{Space-time} $:=(\overline{L}_{n},g)$-space. If necessary, $n$ could be
specified to $n=3,4,$ etc.

\textit{Space} $:=$ $n-1$ sub manifold of a $(\overline{L}_{n},g)$-space.

\textit{Time} $:=$ $1$ dimensional sub manifold of a $(\overline{L}_{n}%
,g)$-space, orthogonal to a given $n-1$ sub manifold of a $(\overline{L}%
_{n},g)$-space, interpreted as a space.

\textit{Space-time distance = distance in a space-time} $:=$ the line-element
(the metric element) $ds=\pm\mid g(d,d)\mid^{1/2}~$in a space-time with
\cite{Manoff-8c}
\[
ds^{2}=\pm\mid ds\mid^{2}=g(d,d)=g_{\overline{i}\overline{j}}\cdot d^{i}\cdot
d^{j}=\pm\mid g(d,d)\mid\text{ \ \ ,}%
\]
where $d=d^{i}\cdot e_{i}=dx^{i}\cdot\partial_{i}$ is the ordinary
differential over a $(\overline{L}_{n},g)$-space.

\textit{Space distance = distance in a space} $:=$ the line-element
$dl=\pm\mid g(d_{\perp},d_{\perp})\mid$ in the space \ with \cite{Manoff-8c}%
\[
dl^{2}=\pm\mid dl\mid^{2}=g(d_{\perp},d_{\perp})=g_{\overline{i}\overline{j}%
}\cdot d_{\perp}^{i}\cdot d_{\perp}^{j}=\pm\mid g(d_{\perp},d_{\perp}%
)\mid\text{ ,}%
\]
where $d_{\perp}=\overline{g}[h_{u}(d)]=l_{d_{\perp}}\cdot n_{\perp}$,
$g(n_{\perp},n_{\perp})=\pm1$, $\overline{g}=g^{ij}\cdot\partial_{i}%
.\partial_{j}$ is the contravariant metric in the $(\overline{L}_{n}%
,g)$-space, $h_{u}=g-\frac{1}{g(u,u)}\cdot g(u)\otimes g(u)$ is the projective
metric, $u$ is a contravariant\ non-null (non-isotropic) vector field,
orthogonal to the given space, $g(u,u)=\pm l_{u}^{2}\neq0$, $h_{u}(u)=0$.

\textit{Time distance = distance in time} $:=$ the line-element $d\tau=\pm\mid
g(d_{\shortparallel},d_{\shortparallel})\mid$ \ in the time with%
\[
d\tau^{2}=\pm\mid d\tau\mid^{2}=\frac{1}{l_{u}^{2}}\cdot g(d_{\shortparallel
},d_{\shortparallel})=\frac{1}{l_{u}^{2}}\cdot g_{\overline{i}\overline{j}%
}\cdot d_{\shortparallel}^{i}\cdot d_{\shortparallel}^{j}=\pm\frac{1}%
{l_{u}^{2}}\cdot\mid g(d_{\shortparallel},d_{\shortparallel})\mid\text{ ,}%
\]
where $d_{\shortparallel}=\frac{1}{g(u,u)}\cdot g(u,d)\cdot u=d\tau\cdot
u=d\tau\cdot l_{u}\cdot n_{\shortparallel}$, $g(d_{\perp},d_{\shortparallel
})=0$. The vector field $u=\frac{d}{d\tau}=l_{u}\cdot n_{\shortparallel}$,
$g(n_{\shortparallel},n_{\shortparallel})=\pm1$, $g(n_{\perp}%
,n_{\shortparallel})=0$, is a contravariant\ non-null (non-isotropic) vector
field, orthogonal to the given space and tangential to a congruence of lines
(non-intersecting curves), orthogonal to the space.

The vector field $u$ is interpreted as a tangent vector field to a congruence
of lines considered as world lines of material points (mass elements,
observers, emitters, etc.) moving in space-time.

A \textit{world line} is a line with a tangent vector $u$ orthogonal to its
corresponding space.

Every space-time distance $ds$ with $ds^{2}$ could be represented by means of
the space distance and the time distance in the form%
\[
ds^{2}=g(d,d)=\pm l_{u}^{2}\cdot d\tau^{2}\mp dl^{2}\text{ \ \ .}%
\]

The space distance $dl$ is the distance between two points lying in the space
orthogonal to the vector field $u$.

The space distance $dl$ and the time distance $d\tau$ could be related by the
introduction of the notion of the absolute value of the relative velocity
$l_{v}$
\[
l_{v}=\frac{dl}{d\tau}%
\]
between two points lying in one and the same space.

Every space distance is measured on the basis of an introduced measure unit
for distance (length) such as meter, inch, yards etc.

Time is the measure of \ a limited or periodical process. The measure unit for
the time $T_{0}$ is introduced by mankind in different ways. Usually, the
measure unit is related to a periodical process. The periodical process is,
from its side, connected to the notion of a clock. If a clock is moving with
an observer it is called standard clock of the observer. The duration of a
process could be measured by different observers with different measure units
for the time distance.

3. Every space-time could be \ decomposed (at least locally) in a pair (space,
time). The decomposition is identical with the $(n-1)+1$ projection of a
space-time \cite{Manoff-1}. This operation is not unique. \ There is a set of
pairs corresponding to a given space-time. At the same time, the space-time
interval $ds$ could be decomposed in corresponding to the pairs forms. If we
now consider two pairs (space, time) we could express the square of the
space-time interval $ds^{2}$ in two different forms%
\[
ds^{2}=\pm l_{us}^{2}\cdot d\tau_{s}^{2}\mp dl_{s}^{2}=\pm l_{u}^{2}\cdot
d\tau^{2}\mp dl^{2}\text{ \ .}%
\]

Therefore, the invariance of the form of the space-time interval, considered
from two different pairs (space, time) leads to the relations%
\[
\frac{d\tau^{2}}{d\tau_{s}^{2}}=\frac{l_{us}^{2}-l_{vs}^{2}}{l_{u}^{2}%
-l_{v}^{2}}\text{ \ \ , \ \ \ }l_{vs}^{2}=\frac{dl_{s}^{2}}{d\tau^{2}}\text{
\ , \ \ \ \ \ \ \ }l_{v}^{2}=\frac{dl^{2}}{d\tau^{2}}\text{ \ .}%
\]

If $l_{vs}=0$ then there is a proportionality between the space-time interval
$ds$ and the time interval $d\tau_{s}$%
\[
ds^{2}=\pm l_{us}^{2}\cdot d\tau_{s}^{2}\text{ \ \ \ . }%
\]

The same conclusion is valid for the other pair of (space, time) if $l_{v}=0$%
\[
ds^{2}=\pm l_{u}^{2}\cdot d\tau^{2}\text{ \ \ \ .}%
\]

This proportionality is usually used for identification of the space-time
interval $ds$ with the time interval $d\tau$ under the additional assumption
that $l_{u}=l_{us}=const.$ There is neither a mathematical nor a physical
reason for the last (above) additional assumption. Nevertheless, the condition
$l_{u}=l_{us}=const.$ is a major assumption in general relativity (on the
analogy of special relativity), where $l_{u}$ is interpreted as the absolute
value of the velocity of a light signal \ in vacuum. In general relativity the
absolute value of a light signal is normalized to $l_{u}=c=const.$, or
$l_{u}=1$. Then $g(u,a)=g(u,a_{\parallel})=0$. The acceleration $a=a_{\perp
}=\overline{g}[h_{u}(a)]$ is orthogonal to the vector field $u$. It is lying
in the sub space orthogonal to $u$. There are two reasons for the
normalization of $u$: one is from mathematical point of view, and the other is
from physical point of view. From mathematical point of view, every non-null
(non-isotropic) contravariant vector field $u$ could be normalized by the use
of the absolute value $l_{u}$ of its length in the form
\begin{align*}
\overline{u}  &  =\pm\cdot\frac{c_{0}}{l_{u}}\cdot u\text{ \thinspace
\thinspace\thinspace, \thinspace\thinspace\thinspace\thinspace\thinspace
\thinspace\thinspace\thinspace\thinspace}g(\overline{u},\overline{u}%
)=\frac{c_{0}^{2}}{l_{u}^{2}}\cdot g(u,u)=\pm c_{0}^{2}\text{ \thinspace
\thinspace,}\\
c_{0}  &  =\text{const. }\neq0\text{ .}%
\end{align*}

From physical point of view, \textit{it is assumed} that a light signal is
propagating with constant absolute value $l_{u}=~const$. from point of view of
the frame of reference of an observer. This point of view leads to its
mathematical realization by means of the normalization of the vector field
$u$. The existing experimental facts about the constancy of the speed of light
cannot cover the whole range of possible decompositions of space-time in pairs
(space, time) related to different frames of reference of different observers
(s. the considerations below). Moreover, the existence of the proportionality
between space-time intervals and time intervals is not related to a special
choice of the velocity parameter $l_{u}$ (respectively $l_{us})$. One can
chose signals of different types (light signals, sound \ signals, periodic
emission of particles etc.) with different velocities to define a time
distance $d\tau$ as well as the relation between it and the space-time
distance $ds$.

A \textit{frame of reference} is determined by the set of three geometric
objects \cite{Manoff-4}:

\begin{itemize}
\item A non-null (time like if $dim\,M=4$) contravariant vector field $u\in
T(M)$.

\item A tangent sub space $T_{x}^{\perp u}(M)$ orthogonal to $u$ at every
point $x\in M$, where $u$ is defined.

\item (Contravariant) affine connection $\nabla=\Gamma$. It determines the
type of transport along the trajectory to which $u$ is a tangent vector field.
$\Gamma$ is related to the covariant differential operator $\nabla_{u}$ along
$u$ \cite{Manoff-1}.

\textit{Remark}. The result $\nabla_{u}T$ of the action of $\nabla_{u}$ on a
tensor field $T\in\otimes^{k}~_{l}(M)$ is called covariant derivative of $T$
along $u$.
\end{itemize}

Then the definition of a frame of reference reads \cite{Manoff-4}

The set $FR\sim\lbrack u,T^{\perp u}(M),\nabla=\Gamma,\nabla_{u}]$ is called
frame of reference in a differentiable manifold $M$\ considered as a model of
the space or of the space-time.

The notion of \textit{periodic signal} could be defined from physical and from
mathematical point of view.

(i) From physical point of view a periodic signal in a $(\overline{L}_{n}%
,g)$-space, considered as a model of space-time, is characterized by:

- A periodic process, characterized by its direction and\ frequency,
transferred by an emitter and received by an observer (detector).

- A periodic process with finite velocity of propagation from point of view of
the observer, characterized by its absolute value of the velocity of propagation.

(ii) From mathematical point of view a periodic signal in a $(\overline{L}%
_{n},g)$-space, considered as a model of space-time, is characterized by:

\begin{itemize}
\item Isotropic (null) contravariant vector field $\widetilde{k}%
:g(\widetilde{k},\widetilde{k})=0$, $\widetilde{k}\in T(M)$, $dim\,M=n$,
$sgn\,g=n-2$ or $sgn\,g=-n+2$, determining the direction of the propagation of
a periodic signal \textit{in space-time}. $M$ is the differentiable manifold
with dimension $n$, provided with affine connections and metrics, $T(M)$ is
the tangent space over $M:T(M)=\cup_{x\in M}T_{x}(M)$.

\item Non-isotropic contravariant vector field $u:g(u,u)=e=\pm l_{u}^{2}\neq
0$.\thinspace The vector field $u\in T(M)$ is interpreted as the
\textit{velocity vector field of an observer (detector)}.

\item $l_{u}^{2}=\pm g(u,u)>0$, interpreted as the \textit{absolute value of
the velocity of a periodic signal} with respect to the proper frame of
reference of an observer. The sign before g(u,u) is depending on the signature
of the metric of the space-time.

\item Scalar product of $\widetilde{k}$ and $u:g(\widetilde{k},u)=\omega>0$,
interpreted as the \textit{frequency of a periodic signal} with respect to the
proper frame of reference of an observer (detector).

\item The \textit{space direction of the propagation of a periodic signal} is
given by the contravariant vector field $k_{\perp}$%
\begin{equation}
k_{\perp}=\overline{g}[h_{u}(\widetilde{k})]=g^{ij}\cdot h_{\overline
{j}\overline{k}}\cdot\widetilde{k}\,^{k}\cdot\partial_{i}\text{ \thinspace
\thinspace\thinspace\thinspace\thinspace\thinspace\thinspace,} \label{1.4}%
\end{equation}
where
\begin{align*}
g(\widetilde{k},\widetilde{k})  &  =0\text{ \thinspace\thinspace
\thinspace\thinspace\thinspace\thinspace\thinspace,\thinspace\thinspace
\thinspace\thinspace\thinspace\thinspace\thinspace\thinspace\thinspace
\thinspace\thinspace\thinspace\thinspace\thinspace\thinspace\thinspace
}\overline{g}=g^{ij}\cdot\partial_{i}.\partial_{j}\,\,\,\,\,\,\,\text{,}\\
\partial_{i}.\partial_{j}\,\,  &  =\frac{1}{2}\cdot(\partial_{i}%
\otimes\partial_{j}+\partial_{j}\otimes\partial_{i})\,\,\text{\thinspace
\thinspace\thinspace\thinspace\thinspace\thinspace\thinspace\thinspace,}\\
h_{u}  &  =g-\frac{1}{g(u,u)}\cdot g(u)\otimes g(u)\text{ \thinspace
\thinspace\thinspace\thinspace,}\\
g(u,u)  &  =\pm l_{u}^{2}=e\neq0\,\,\,\,\,\,\,\text{.}%
\end{align*}

\end{itemize}

The frequency $\omega$ of the periodic standard signal (the periodic signal
sent by a standard emitter) is determined by the relations \cite{Manoff-8c}:
\begin{equation}
\omega=g(u,\widetilde{k})=l_{u}\cdot g(\widetilde{n}_{\perp},k_{\perp
})\,\,\,\,\,\,\text{,} \label{1.5}%
\end{equation}
where
\begin{equation}
k_{\perp}=\mp l_{k_{\perp}}\cdot\widetilde{n}_{\perp}=\mp\frac{\omega}{l_{u}%
}\cdot\widetilde{n}_{\perp}\text{ \thinspace\thinspace\thinspace
\thinspace,\thinspace\thinspace\thinspace\thinspace\thinspace\thinspace
\thinspace\thinspace\thinspace\thinspace\thinspace}g(\widetilde{n}_{\perp
},\widetilde{n}_{\perp})=\mp1\text{ \thinspace\thinspace\thinspace.}
\label{1.6}%
\end{equation}

\begin{itemize}
\item By the use of the relations
\begin{align}
l_{u}  &  =\lambda\cdot\nu=\lambda\cdot\frac{\omega}{2\cdot\pi}%
\,\,\,\,\,\,\,\,\text{,}\label{2.4}\\
\omega &  =2\cdot\pi\cdot\frac{l_{u}}{\lambda}=l_{u}\cdot g(\widetilde
{n}_{\perp},k_{\perp})\,\,\,\,\,\,\,\text{,} \label{2.5}%
\end{align}
the expression for the length $\lambda$, corresponding to the frequency
$\omega~$ of the periodic signal, could be found in the form
\begin{equation}
\lambda=\frac{2\cdot\pi}{g(\widetilde{n}_{\perp},k_{\perp})}%
\,\,\,\,\,\,\,\,\,\,\text{,} \label{2.6}%
\end{equation}
and therefore,
\begin{equation}
g(\widetilde{n}_{\perp},k_{\perp})=\frac{2\cdot\pi}{\lambda}%
\,\,\,\,\,\,\text{.} \label{2.7}%
\end{equation}

\end{itemize}

The projection of $k_{\perp}$ on its unit vector $\widetilde{n}_{\perp}$ has
exact relation to the length $\lambda$ of the periodic signal with frequency
$\omega$.

\section{Standard periodic emitter (clock)}

1. A \textit{standard emitter} is an emitter moving together with an observer
(detector) in space-time and lying at rest in the proper frame of reference of
the observer (detector). The proper frame of reference of the standard emitter
could be identified with the proper frame of reference \cite{Manoff-2} of the
observer (detector).

The notion of a clock is closely related to the notion of a periodic signal.
\textquotedblright A clock is a physical device consisting of an oscillator
running at some angular frequency $\omega$ and a counter that counts the
cycles. The period of the oscillator, $T=2\pi/\omega$, is calibrated in some
standard oscillator. The counter simply counts the cycles of the oscillator.
Since some epoch, or the event at which the count started, we say that a
quantity of time equal to $NT$ has elapsed, if $N$ cycles have been
counted\textquotedblright\ \cite{Bahder}.

A \textit{standard clock} is a clock moving with an observer (detector) and
with his standard emitter.

\subsection{Variation of the velocity of a periodic signal along the world
line of a standard emitter}

Let us now consider the motion of a standard emitter moving with an observer
in a $(\overline{L}_{n},g)$-space considered as a model of space-time.

1. Let $\tau$ be the proper time (parameter) of the world line (trajectory)
$x^{i}(\tau)$, $i=1,...,n$, $dim\,M=n$, $n=4$, in a space-time of a standard
emitter [i.e. of an emitter lying at rest with an observer (detector)]. The
proper time can be introduced by the use of a calibrated standard emitter
(clock). The tangent vector field along the world line of the standard emitter
could be written in the form
\begin{equation}
u=\frac{d}{d\tau}=\frac{dx^{i}}{d\tau}\cdot\partial_{i}=u^{i}\cdot\partial
_{i}\text{ \thinspace\thinspace\thinspace\thinspace,\thinspace\thinspace
\thinspace\thinspace\thinspace\thinspace\thinspace\thinspace\thinspace
\thinspace\thinspace\thinspace}u^{i}=\frac{dx^{i}}{d\tau}%
\,\,\,\,\,\,\,\text{.} \label{4.1}%
\end{equation}

Its absolute length $l_{u}$ is defined by the expression \cite{Manoff-1}
\begin{equation}
\pm l_{u}^{2}=g(u,u)=g_{\overline{i}\overline{j}}\cdot u^{i}\cdot
u^{j}\,\,\,\,\,\text{,\thinspace\thinspace\thinspace\thinspace\thinspace
\thinspace\thinspace\thinspace\thinspace\thinspace\thinspace\thinspace
\thinspace\thinspace\thinspace\thinspace\thinspace}g_{\overline{i}\overline
{j}}=f^{k}\,_{i}\cdot f^{l}\,_{j}\cdot g_{kl}\text{ \thinspace\thinspace
\thinspace\thinspace,} \label{4.2}%
\end{equation}

2. The variation of the absolute value $l_{u}$ of the velocity of a periodic
signal of a standard emitter could be found \ by the use of the relations
\begin{align}
\nabla_{u}(l_{u}^{2})  &  =u(l_{u}^{2})=2\cdot l_{u}\cdot u(l_{u})=2\cdot
l_{u}\cdot\frac{dl_{u}}{d\tau}=\nonumber\\
&  =\pm\nabla_{u}[g(u,u)]=\pm\lbrack(\nabla_{u}g)(u,u)+2\cdot g(u,a)]\text{
\thinspace\thinspace\thinspace,}\label{4.7}\\
a  &  =\nabla_{u}u\text{ \thinspace\thinspace\thinspace\thinspace.}\nonumber
\end{align}
in the form
\begin{equation}
l_{u}\cdot\frac{dl_{u}}{d\tau}=\pm\lbrack g(u,a)+\frac{1}{2}\cdot(\nabla
_{u}g)(u,u)]\,\,\,\,\,\,\,\text{.} \label{4.8}%
\end{equation}

3. If we consider the conditions under which the absolute value $l_{u}$ of the
velocity of a periodic signal appears as a conserved quantity in a
$(\overline{L}_{n},g)$-space, we can prove the following propositions:

\textit{Proposition 1}. The necessary and sufficient condition for the
absolute value $l_{u}$ of the velocity of a periodic signal to be a conserved
quantity along the world line of a standard emitter with tangent vector field
$u$ is the condition
\begin{equation}
g(u,a)=-\frac{1}{2}\cdot(\nabla_{u}g)(u,u)\,\,\,\,\,\,\,\,\,\text{,}
\label{4.9}%
\end{equation}
equivalent to the condition
\begin{equation}
(\pounds _{u}g)(u,u)=0\text{ .} \label{4.10}%
\end{equation}

The proof is trivial. One should only take into account the relations
\begin{align*}
l_{u}\cdot\frac{dl_{u}}{d\tau}  &  =\pm\lbrack g(u,a)+\frac{1}{2}\cdot
(\nabla_{u}g)(u,u)]\,\,\text{\thinspace\thinspace\thinspace\thinspace
\thinspace,\thinspace\thinspace\thinspace\thinspace\thinspace\thinspace
\thinspace\thinspace\thinspace\thinspace\thinspace}l_{u}\neq0\text{
\thinspace\thinspace,}\\
\pounds _{u}[g(u,u)]  &  =u[g(u,u)]=\nabla_{u}[g(u,u)]=(\pounds _{u}%
g)(u,u)\text{ \thinspace\thinspace.}%
\end{align*}

\textit{Proposition 2}. A sufficient condition for the absolute value
$l_{u}\neq0$ of the velocity of a periodic signal of an emitter to be a
constant quantity ($l_{u}=const.\neq0$) along the world line of the emitter
with tangent vector field $u$ is the condition
\begin{equation}
\pounds _{u}g=0\,\,\,\,\,\,\,\text{,} \label{4.11}%
\end{equation}
i.e. if the vector field $u$ is tangent vector field to the world line of a
standard emitter then the absolute value $l_{u}$ of the velocity of its
periodic signals is a constant quantity ($l_{u}=const.\neq0$) along the world
line of the emitter when $u$ is a Killing vector field.

The proof is trivial.

\subsection{Variation of the frequency of a periodic signal along the world
line of an emitter}

If an emitter is used as a standard emitter by an observer on his world line
for comparison of the incoming periodic signals from another emitters then the
variation of the frequency of the standard emitter is of great importance for
the correct determination of the frequency of the incoming signals.

1. From the explicit form of the frequency $\omega$%
\begin{equation}
\omega=l_{u}\cdot g(\widetilde{n}_{\perp},k_{\perp})\,\,\,\,\,\text{,}
\label{2.1}%
\end{equation}
where \cite{Manoff-8c}
\begin{equation}
k_{\perp}=\mp\frac{\omega}{l_{u}}\cdot\widetilde{n}_{\perp}%
\,\,\,\,\,\,\,\text{,\thinspace\thinspace\thinspace\thinspace\thinspace
\thinspace\thinspace\thinspace\thinspace\thinspace\thinspace\thinspace
\thinspace}g(\widetilde{n}_{\perp},u)=0\text{\thinspace\thinspace
\thinspace\thinspace\thinspace\thinspace,} \label{2.2}%
\end{equation}
it follows that
\begin{align}
\nabla_{u}\omega &  =u\omega=\frac{d\omega}{d\tau}=\nabla_{u}[l_{u}\cdot
g(\widetilde{n}_{\perp},k_{\perp})\,]=\nonumber\\
&  =ul_{u}\cdot g(\widetilde{n}_{\perp},k_{\perp})+\nonumber\\
&  \,+l_{u}\cdot\lbrack(\nabla_{u}g)(\widetilde{n}_{\perp},k_{\perp
})\,+g(\nabla_{u}\widetilde{n}_{\perp},k_{\perp})+g(\widetilde{n}_{\perp
},\nabla_{u}k_{\perp})]\text{ \thinspace\thinspace.} \label{2.3}%
\end{align}

By the \ use of the conditions $\pounds _{u}k_{\perp}=0$,\thinspace
\thinspace\thinspace\thinspace$g(u,k_{\perp})=0\,$, (the relations are the
necessary and sufficient conditions for the existence of co-ordinates curves
to which $u$ and $k_{\perp}$ appear as tangent vectors at every point of the
corresponding curve \cite{Bishop}), $\ $the equation for the frequency
$\omega$ follows in the form%
\[
\frac{d}{d\tau}(log\frac{\omega}{l_{u}})=\mp\lbrack d(\widetilde{n}_{\perp
},\widetilde{n}_{\perp})+\frac{1}{2}\cdot(\nabla_{u}g)(\widetilde{n}_{\perp
},\widetilde{n}_{\perp})]\,\,\,\,\,\,\,\text{,}%
\]

where $d$ is the deformation velocity tensor \cite{Stephani}, \cite{Manoff-1}
\begin{equation}
d=\sigma+\omega+\frac{1}{n-1}\cdot\theta\cdot h_{u}\text{ \thinspace
\thinspace\thinspace.} \label{2.9}%
\end{equation}

The trace free covariant symmetric tensor $\sigma$ is the shear velocity
tensor. The antisymmetric covariant tensor $\omega$ (not necessary for the
further investigations) is the rotation velocity tensor. The invariant
$\theta$ is the expansion velocity invariant \cite{Stephani}, \cite{Manoff-1}.

On the other side,
\begin{align}
d(\widetilde{n}_{\perp},\widetilde{n}_{\perp})  &  =(\sigma+\omega+\frac
{1}{n-1}\cdot\theta\cdot h_{u})(\widetilde{n}_{\perp},\widetilde{n}_{\perp
})=\nonumber\\
&  =\mp\lbrack\frac{1}{n-1}\cdot\theta\mp\sigma(\widetilde{n}_{\perp
},\widetilde{n}_{\perp})]\,\,\,\,\,\,\text{,} \label{2.13}%
\end{align}%
\begin{equation}
\frac{1}{n-1}\cdot\theta\mp\sigma(\widetilde{n}_{\perp},\widetilde{n}_{\perp
})=H\,\,\,\,\,\,\,\,\,\,\,\,\text{,\thinspace\thinspace\thinspace
\thinspace\thinspace\thinspace\thinspace\thinspace\thinspace\thinspace
\thinspace\thinspace\thinspace\thinspace}d(\widetilde{n}_{\perp},\widetilde
{n}_{\perp})=\mp H\,\,\,\,\,\,\text{.\thinspace\thinspace} \label{2.14}%
\end{equation}

The function $H=H(\tau)$ is the s.c. Hubble function \cite{Manoff-8c}.
Therefore, we can write now the equation for the frequency $\omega$
\begin{equation}
\frac{d}{d\tau}(log\frac{\omega}{l_{u}})=H\mp\frac{1}{2}\cdot(\nabla
_{u}g)(\widetilde{n}_{\perp},\widetilde{n}_{\perp})\text{ \thinspace
\thinspace\thinspace\thinspace\thinspace.} \label{2.15}%
\end{equation}

Its solution follows in the form
\begin{align*}
\frac{\omega}{l_{u}}  &  =\frac{\omega_{0}}{l_{u0}}\cdot exp(\int[H\mp\frac
{1}{2}\cdot(\nabla_{u}g)(\widetilde{n}_{\perp},\widetilde{n}_{\perp})]\cdot
d\tau)\text{ \thinspace\thinspace\thinspace\thinspace\thinspace,}\\
\omega_{0}  &  =\text{ const.,\thinspace\thinspace\thinspace\thinspace
\thinspace\thinspace\thinspace\thinspace\thinspace\thinspace\thinspace
\thinspace\thinspace\thinspace\thinspace}l_{u0}=\text{ const.}%
\end{align*}

2. By the use of the relations
\begin{align}
l_{u}  &  =\lambda\cdot\nu=\lambda\cdot\frac{\omega}{2\cdot\pi}%
\,\,\,\,\,\,\,\,\text{,}\label{2.19a}\\
\omega &  =2\cdot\pi\cdot\frac{l_{u}}{\lambda}=l_{u}\cdot g(\widetilde
{n}_{\perp},k_{\perp})\,\,\,\,\,\,\,\text{,} \label{2.19b}%
\end{align}

the expression for the length $\lambda$ follows in the form%
\begin{equation}
\lambda=\lambda_{0}\cdot exp\{-\int[H\mp\frac{1}{2}\cdot(\nabla_{u}%
g)(\widetilde{n}_{\perp},\widetilde{n}_{\perp})]\cdot d\tau
\}\,\,\,\,\,\text{.} \label{2.22}%
\end{equation}

Therefore, the length $\lambda$ of the standard periodic signal will decrease
in the proper frame of the standard emitter (observer) if
\begin{equation}
H\mp\frac{1}{2}\cdot(\nabla_{u}g)(\widetilde{n}_{\perp},\widetilde{n}_{\perp
})>0\,\,\,\text{,\thinspace\thinspace\thinspace\thinspace\thinspace
\thinspace\thinspace\thinspace\thinspace\thinspace}\lambda<\lambda_{0}\text{
\thinspace\thinspace\thinspace\thinspace,} \label{2.23}%
\end{equation}
and $\lambda$ will increase when
\begin{equation}
H\mp\frac{1}{2}\cdot(\nabla_{u}g)(\widetilde{n}_{\perp},\widetilde{n}_{\perp
})<0\,\,\,\,\,\,\,\,\text{,\thinspace\thinspace\thinspace\thinspace
\thinspace\thinspace\thinspace\thinspace\thinspace}\lambda>\lambda
_{0}\,\,\,\,\,\,\text{.} \label{2.24}%
\end{equation}

If
\begin{equation}
H\mp\frac{1}{2}\cdot(\nabla_{u}g)(\widetilde{n}_{\perp},\widetilde{n}_{\perp
})=0\,\,\,\,\,\,\,\,\text{,} \label{2.25}%
\end{equation}
then
\[
\lambda=\lambda_{0}=\text{ const.}%
\]

3. From the equation for $l_{u}$%
\begin{equation}
\frac{1}{2}\cdot\frac{dl_{u}^{2}}{d\tau}=\pm\lbrack g(u,a)+\frac{1}{2}%
\cdot(\nabla_{u}g)(u,u)] \label{2.26}%
\end{equation}
we can find the expression for $l_{u}$%
\[
\frac{dl_{u}^{2}}{d\tau}=\pm2\cdot\lbrack g(u,a)+\frac{1}{2}\cdot(\nabla
_{u}g)(u,u)]\text{ \thinspace\thinspace\thinspace\thinspace\thinspace
\thinspace\thinspace\thinspace.}%
\]

If we now substitute the expression for $l_{u}$ in the expression for $\omega$
we can find the general relation for the variation of the frequency $\omega
$~under the variation of the absolute value $l_{u}$ of a standard periodic
signal with the proper time $\tau$ of the emitter
\begin{align}
\omega &  =\omega_{0}\cdot(1\pm\frac{2}{l_{u0}^{2}}\cdot\int[g(u,a)+\frac
{1}{2}\cdot(\nabla_{u}g)(u,u)]\cdot d\tau)^{1/2}\cdot\nonumber\\
&  \cdot exp(\int[H\mp\frac{1}{2}\cdot(\nabla_{u}g)(\widetilde{n}_{\perp
},\widetilde{n}_{\perp})]\cdot d\tau)\,\,\,\,\,\,\text{.} \label{2.28}%
\end{align}

As a final results, for the variations of the absolute value $l_{u}$, the
frequency $\omega$, and the length $\lambda$ of a standard periodic signal
with the proper time $\tau$ of a standard emitter we obtain the relations
\begin{align}
l_{u}  &  =(l_{u0}^{2}\pm2\cdot\int[g(u,a)+\frac{1}{2}\cdot(\nabla
_{u}g)(u,u)]\cdot d\tau)^{1/2}\,\,\,\,\,\,\,\,\,\,\text{,\thinspace
\thinspace\thinspace\thinspace\thinspace\thinspace\thinspace\thinspace
\thinspace}\tag{A}\\
\text{\thinspace\thinspace\thinspace\thinspace\thinspace}l_{u0}^{2}  &
=\,\text{const.\thinspace}>0\text{ \thinspace,}\nonumber
\end{align}%
\begin{align}
\omega &  =\omega_{0}\cdot(1\pm\frac{2}{l_{u0}^{2}}\cdot\int[g(u,a)+\frac
{1}{2}\cdot(\nabla_{u}g)(u,u)]\cdot d\tau)^{1/2}\cdot\nonumber\\
&  \cdot exp(\int[H\mp\frac{1}{2}\cdot(\nabla_{u}g)(\widetilde{n}_{\perp
},\widetilde{n}_{\perp})]\cdot d\tau)\,\,\,\,\,\,\,\,\text{,} \tag{B}%
\end{align}%
\begin{align}
\lambda &  =\lambda_{0}\cdot exp\{-\int[H\mp\frac{1}{2}\cdot(\nabla
_{u}g)(\widetilde{n}_{\perp},\widetilde{n}_{\perp})]\cdot d\tau
\}\,\,\,\,\,\,\text{,}\tag{C}\\
\lambda_{0}  &  =\text{ const.}\nonumber
\end{align}

\subsection{Variation of the velocity and the frequency of periodic signals of
a standard oscillator (clock)}

\subsubsection{Variation of the period of a standard clock}

Let us now consider the change of the frequency of periodic signals of a
standard oscillator used as a standard clock in the frame of reference of an
observer. The period of the oscillator, $T=2\cdot\pi/\omega$, is calibrated in
some standard oscillator. If the frequency $\omega$ is changing along the
world line of the oscillator then the period $T$ is also changing in the
corresponding form
\begin{align}
T  &  =\frac{2\cdot\pi}{\omega}=T_{0}\cdot(1\pm\frac{2}{l_{u0}^{2}}\cdot
\int[g(u,a)+\frac{1}{2}\cdot(\nabla_{u}g)(u,u)]\cdot d\tau)^{-1/2}%
\cdot\nonumber\\
&  \cdot exp(-\int[H\mp\frac{1}{2}\cdot(\nabla_{u}g)(\widetilde{n}_{\perp
},\widetilde{n}_{\perp})]\cdot d\tau)\,\,\,\,\,\,\,\,\text{,}\label{3.1}\\
T_{0}  &  =\frac{2\cdot\pi}{\omega_{0}}=\text{ const.}\nonumber
\end{align}

Therefore, a standard oscillator (clock) would not change its frequency and
period if and only if the following conditions are fulfilled
\begin{align}
g(u,a)+\frac{1}{2}\cdot(\nabla_{u}g)(u,u)  &  =0\,\,\,\,\,\,\,\text{,}%
\label{3.2}\\
H\mp\frac{1}{2}\cdot(\nabla_{u}g)(\widetilde{n}_{\perp},\widetilde{n}_{\perp
})  &  =0\,\,\,\,\,\,\text{.} \label{3.3}%
\end{align}

These conditions are in generally not fulfilled even in the Einstein theory of
gravitation, where the above relations should also be valid in their special
forms
\begin{align}
g(u,a)  &  =0\,\,\,\,\,\,\,\text{,}\label{3.4}\\
H  &  =0\,\,\,\,\,\,\text{.} \label{3.5}%
\end{align}

In all cosmological models in general relativity with Hubble function
$H=H(\tau)$ different from zero and $g(u,a)=0$ a standard clock will move in
space-time with a period $T$ obeying the condition
\begin{equation}
T=\frac{2\cdot\pi}{\omega}=T_{0}\cdot exp(-\int H\cdot d\tau
)\,\,\,\,\,\,\,\,\,\,\text{.} \label{3.6}%
\end{equation}

Furthermore, the condition $g(u,a)=0$ is considered as a corollary of the
assumption of the constant value $l_{u}=c=const.$ or $l_{u}=1$ of the speed of
light. There is no unique physical argument for the last assumption if a
theory of gravitation has to describe the behavior of physical systems
including the speed of propagation of their interactions on the basis of its
own structures.

\section{Calibrated standard emitter (clock)}

If a standard emitter (clock) changes its frequency and the absolute value of
the velocity of its signals along the world line of the observer the question
arises how these changes could be registered and compared to the changes of
other standard emitters. The emitter (clock) to which the data of the standard
clock should be compared should be again a standard emitter but with constant
frequency and constant absolute value of the velocity of its signals along the
world line of the observer.

A standard emitter \ (clock) with constant frequency and constant absolute
value of the velocity of its signals along the world line of the observer is
called a \textit{calibrated standard emitter (clock)}.

Because of its properties, a calibrated standard emitter (clock) should be
transported along the world line of the observer by a transport, different
from the covariant transports of the standard emitters along the world line of
the observer. This type of transport should preserve the signals'
characteristics of the emitted signals (frequency, length, absolute value of
the velocity of the emitted signals). Therefore, we should find such type of
transports along the world line of the observer. From mathematical point of
view these transports should preserve the length of the vector field $u$
(related to the absolute value of the velocity of the signals) as well as the
length of the vectors, orthogonal to $u$ (related to the frequency and the
length of the signals).

Transports preserving the lengths of vector fields, orthogonal to each other
in a $(\overline{L}_{n},g)$-space, are called \textit{generalized Fermi-Walker
transports} (FWT). The theory of FWT is worked out in details in
\cite{Manoff-1}. Brief reviews of the properties of FWT are given in
\cite{Manoff-4a} and in \cite{Manoff-4b}.

In general, if the transport along the world line is determined by a
generalized type of a Fermi-Walker transport \cite{Manoff-1}, \cite{Manoff-4a}%
, \cite{Manoff-4b} the length of the vector field $u$ remains a constant
quantity along its trajectory as well as the length of the vectors orthogonal
to it. In this case, the following conditions are fulfilled%
\[
l_{u}=l_{u0}=const.\neq0\text{ \ \ ,}%
\]%
\[
k_{\perp}=\mp\frac{\omega}{l_{u}}\cdot\widetilde{n}_{\perp}=\mp l_{k_{\perp}%
}\cdot\widetilde{n}_{\perp}\text{ \ \ \ \ ,}%
\]%
\[
l_{k_{\perp}}=const.=\frac{\omega}{l_{u}}\text{ \ \ .}%
\]

Since $l_{u}=const.\neq0$, it follows that%
\[
\omega=\omega_{0}=const.\text{, \ \ \ \ \ \ \ \ \ }\lambda=\lambda_{0}=const.
\]

\ A calibrated standard clock could be now used as a device for comparison and
determination of the changes of the frequency and the absolute value of the
velocity of signals, emitted by other (non-calibrated) standard emitters
(clocks). The same conclusion is also valid for non-standard emitters (clocks).

\subsection{Synchronization of calibrated standard clocks}

In the Einstein theory of gravitation the synchronization of calibrated
standard clocks \cite{Tucker}, \cite{Rizzi} means the use of curves on which
the absolute value $l_{u}$ of the velocity of a light signal remains constant.
Such types of curves are the geodesic trajectories of observers in space-time.
But the existence of geodesic trajectories is only a sufficient but not a
necessary condition.

In the general relativity as well as in (pseudo-) Riemannian spaces without
torsion ($V_{n}$-spaces) the necessary and sufficient condition for
$l_{u}=c=const.\neq0$ is the condition $g(u,a)=0$, i.e. the orthogonality
between $u$ and $a=\nabla_{u}u$. On the other side, this condition does not
lead to the condition $l_{u}=c=const.\neq0$ in Weyl's spaces. Therefore, if we
search for a condition for $u$ leading to the condition $l_{u}=c=const.\neq0$
in a $(\overline{L}_{n},g)$-space we should consider other possible preconditions.

For a more precise definition of the notion of synchronization we should
assume that a calibrated standard clock should not change its frequency
$\omega$ and the absolute value $l_{u}$ of the velocity of its periodic
signals if it could be synchronized with other standard clocks.

The notion of \textit{synchronized calibrated standard clocks} could be then
defined in the form:

\textit{Definition}. Two calibrated standard clocks (lying on their world
lines) are called to be synchronized if they do not change their frequency
$\omega$ and the absolute value $l_{u}$ of the velocity of their periodic
signals along their own (proper) \ world lines, i.e. $l_{u}=l_{uo}%
=const.\neq0$ and $\omega=\omega_{0}=const.$ on every world line on which a
standard clock is moving in space-time. This means that the length $l_{u}$ of
the vector field $u$ does not change along its curve to which it is a tangent
vector. In general, if a curve is determined by a generalized type of a
Fermi-Walker transport \cite{Manoff-1}, \cite{Manoff-4a}, \cite{Manoff-4b} the
length of the vector field $u$ remains a constant quantity along its
trajectory as well as the length of the vectors orthogonal to it. In this
case, the following conditions are fulfilled%
\[
l_{u}=const.\neq0\text{ \ \ ,}%
\]%
\[
k_{\perp}=\mp\frac{\omega}{l_{u}}\cdot\widetilde{n}_{\perp}=\mp l_{k_{\perp}%
}\cdot\widetilde{n}_{\perp}\text{ \ \ \ \ ,}%
\]%
\[
l_{k_{\perp}}=const.=\frac{\omega}{l_{u}}\text{ \ \ .}%
\]

Since $l_{u}=const.\neq0$, it follows that%
\[
\omega=\omega_{0}=const.
\]

Therefore, calibrated standard clocks could be synchronized if they are
transported by generalized Fermi-Walker transports along the world lines of
the corresponding observers (detectors).

\section{Conclusions}

In the present paper the variation of the absolute value and the frequency of
a periodic signal sent by a standard emitter is considered. \textit{The
obtained results contradict with the general belief that a standard emitter
does not change the frequency of its periodic signals on its world line
considered also as the world line of an observer (detector) moving together
with the standard emitter}. \ Calibrated standard emitters (clocks) could
exist if they are transported along the world line of the observer by a
generalized Fermi-Walker transport. Calibrated standard emitters (clocks)
should exist if an exact comparison with other standard emitters or with
incoming signals is required. \ The synchronization of calibrated standard
clocks is also possible in spaces with affine connections and metrics under
the condition that these clocks are transported by a generalized Fermi-Walker
transports along their world lines.

\end{document}